\documentclass{ws-mpla}
\usepackage[super]{cite}
\usepackage{graphicx}
\begin{document}


\catchline{}{}{}{}{}

\title{
Schr$\ddot{\rm{o}}$dinger Equation of a particle in an Uniformly Accelerated
Frame and the Possibility of a New kind of Quanta 
}

\author{\footnotesize Sanchari De$^a$}

\address{$^a$Department of Physics, Visva-Bharati, Santiniketan, India 731235\\
E-mail: sancharide19@gmail.com
}

\author{ \footnotesize Sutapa Ghosh$^b$}

\address{$^b$Department of Physics, Barasat Govt. College, 
Barasat 700124, India\\ E-mail:neutronstar12@yahoo.co.in}

\author{Somenath Chakrabarty$^a$}

\address{$^a$Department of Physics, Visva-Bharati, Santiniketan, India 731235\\
E-mail: somenath.chakrabarty@visva-bharati.ac.in
}

\maketitle

\pub{Received (Day Month Year)}{Revised (Day Month Year)}

\begin{abstract}
In this article we have developed a formalism to obtain the Schr$\ddot{\rm{o}}$dinger equation for a particle
in a frame undergoing an uniform acceleration in an otherwise flat
Minkowski space-time geometry. We have presented an exact solution
of the equation and obtained the eigenfunctions and the corresponding
eigenvalues. It has been observed that the Schr$\ddot{\rm{o}}$dinger
equation can be reduced to an one dimensional hydrogen atom problem.
Whereas, the quantized energy levels are exactly identical with
that of an one dimensional quantum harmonic oscillator. Hence considering transitions, we have 
predicted the existence of a new kind of quanta, which will either be emitted or
absorbed if the particles get excited or de-excited respectively.

\end{abstract}

\ccode{PACS Nos.: 
03.65.Ge,03.65.Pm,03.30.+p,04.20.-q 
}

In the case of inertial frame of references, in the relativistic 
scenario, the set of space-time coordinates are
transformed following the well known Lorentz formulas \cite{R1,R2}. However, for a frame undergoing 
an uniform accelerated motion in an otherwise
flat Minkowski space-time geometry, the Lorentz type transformation can 
also be derived to relate the coordinate sets in
accelerated frame with that of inertial set of coordinates \cite{R3,R4,R5}. 
These are the so called Rindler coordinate transformations and are given by: 
\begin{eqnarray}
ct&=&\left (\frac{c^2}{\alpha}+x^\prime\right )\sinh\left (\frac{\alpha t^\prime}
{c}\right ) ~~{\rm{and}}~~ \nonumber \\
x&=&\left (\frac{c^2}{\alpha}+x^\prime\right )\cosh\left (\frac{\alpha t^\prime}
{c}\right ) 
\end{eqnarray}
Hence one can also express the inverse relations
\begin{equation}
ct^\prime=\frac{c^2}{2\alpha}\ln\left (\frac{x+ct}{x-ct}\right )
~~{\rm{and}}~~ x^\prime=(x^2-(ct)^2)^{1/2}-\frac{c^2}{\alpha}
\end{equation}
Where $\alpha$ is the uniform acceleration of the frame.
The Rindler space-time coordinates as shown above (eqns.(1) and (2)) are then just an accelerated frame
transformation of the Minkowski metric of special relativity. 

We know that in the case of Schr$\ddot{\rm{o}}$dinger equation, 
the space time geometry is essentially flat and Euclidean in
nature. On the other hand, in the relativistic picture, in the study of 
quantum mechanical equations,
e.g., Klein-Gordon equation or Dirac equation, the geometry is also flat, 
but Minkowski type. In the curved space-time configuration, where the
geometry is Riemann type, i.e., the formalism 
is based on the general theory of
relativity, a substantial amount of work have already been done on the 
study of both Klein-Gordon equation \cite{R6} and Dirac equation
\cite{R7} (see also \cite{R3,R7a,R7b,R7c}). 
However, so far our knowledge is 
concerned nothing 
has been reported in the literature on the study of
Schr$\ddot{\rm{o}}$dinger equation in uniformly accelerated frame with non-relativistic approximation.

In this article we have developed a formalism to obtain the 
Schr$\ddot{\rm{o}}$dinger equation in an uniformly accelerated frame,
or in otherwise speaking, in accordance with the principle of  equivalence in a rest 
frame in presence of a constant gravitational field. 
Our study is thus based on the principle of equivalence, according to which 
a frame of reference undergoing an
accelerated motion in absence of a gravitational field is equivalent 
to a frame at rest in presence of a gravitational field. To the best
of our knowledge the  study of Schr$\ddot{\rm{o}}$dinger equation
in uniformly accelerated frame has not been done before.

Now it can very easily be shown that under Rindler coordinates
transformation the Minkowski line element changes from \cite{R5}
\begin{equation}
ds^2=d(ct)^2-dx^2-dy^2-dz^2 ~~{\rm{to}}~~ ds^2=\left
(1+\frac{\alpha x^\prime}{c^2}\right)^2d(ct^\prime)^2-{dx^\prime}^2
-{dy^\prime}^2-{dz^\prime}^2
\end{equation}
Now following the formalism of relativistic dynamics of special theory
of relativity, the action integral can be written as \cite{R1}
\begin{equation}
S=-\alpha_0 \int_a^b ds\equiv \int_a^b Ldt
\end{equation}
Then following \cite{R1} after putting $\alpha_0=-m_0 c$, where $m_0$ is 
rest mass of the particle and finally omitting the prime symbols, the
Lagrangian of the particle can be written as \cite{R5}
\begin{equation}
L=-m_0c^2\left [\left ( 1+\frac{\alpha x}{c^2}\right )^2 -\frac{v^2}{c^2}
\right ]^{1/2}
\end{equation}
To develop the non-relativistic version of quantum mechanical formalism for a particle in 
an uniformly accelerated frame, we start with this single particle
classical Lagrangian in Rindler space. 
Here
$v$ the particle velocity along $x$-direction.
Then assuming three dimensional motion, the three momentum vector of the
particle may  be written as
\begin{equation}
\vec p=\frac{m_0\vec v}{\left [ \left (1+\frac{\alpha x}{c^2} \right )^2
-\frac{v^2}{c^2} \right ]^{1/2}}
\end{equation}
Obviously we get back the special theory result for $\alpha=0$.
Hence the classical Hamiltonian of the particle  is given by
\begin{equation}
H=m_0c^2\left ( 1+\frac{\alpha x}{c^2}\right ) \left (
1+\frac{p^2}{m_0^2 c^2}\right )^{1/2}
\end{equation}
Which also reduces to the Hamiltonian of special theory of relativity
for $\alpha=0$.
Now in the quantum mechanical picture, the classical dynamical
variables $x$, $\vec p$ and also $H$ are treated as operators, with
the canonical commutation relations
\begin{equation}
[x,p_x]=i\hbar~~{\rm{and}}~~
[x,p_y]= [x,p_z]=0     
\end{equation}
The Schr$\ddot{\rm{o}}$dinger equation for the particle is then given by
\begin{equation}
H\psi =\left ( 1+\frac{\alpha x}{c^2} \right ) \left ( m_0c^2
+\frac{p^2}{2m_0} \right )\psi =E\psi
\end{equation}
Here the origin of co-ordinate is assumed to be at the centre of
a strongly gravitating object, say a black hole and $x$ is measured along
the positive direction. 
Now considering the eigenvalue form of the operator $1+\alpha
x/c^2$ on
the right hand side, we have
\begin{equation}
\left ( m_0c^2+\frac{p^2}{2m_0}\right )\psi=\frac{1}{1+\frac{\alpha x}{c^2}}E\psi
\end{equation}
Since there is an $x$-dependent term in the right hand side of the above equation, 
the separable form of solution for the Schr$\ddot{\rm{o}}$dinger equation
can obviously be written as 
\begin{equation}
\psi(x,y,z) =NX(x)\exp\left (-\frac{ip_yy}{\hbar}\right ) 
\exp\left (-\frac{ip_zz}{\hbar}\right )
\end{equation}
where $X(x)$ is some unknown function of $x$.
On substituting $\psi(x,y,z)$ in eqn.(10), we have 
\begin{equation}
\left (m_0c^2+\frac{p_y^2}{2m_0}+\frac{p_z^2}{2m_0}\right )X-\frac{\hbar^2}{2m_0}
\frac{d^2X}{dx^2}=\frac{1}{1+\frac{\alpha x}{c^2}}EX
\end{equation}
Now defining the quantity within the first bracket on the left hand side as 
$E_{\perp}$, the orthogonal part of the total
energy, including the rest mass energy of the particle, we have
\begin{equation}
E_{\perp}X-\frac{\hbar^2}{2m_0}\frac{d^2X}{dx^2}=
\frac{1}{1+\frac{\alpha x}{c^2}}EX
\end{equation}
Let us now introduce a new variable $u=1+\alpha x/c^2$ and
define two constant quantities
\[
a=\frac{2m_0c^4}{\hbar^2\alpha^2}E~~{\rm{and}}~~ b=
\frac{8m_0c^4}{\hbar^2\alpha^2}E_\perp,
\] 
then we can express the above equation (eqn.(13)) in
the following form
\begin{equation}
\frac{d^2X}{du^2}+\frac{a}{u}X-\frac{b}{4}X=0
\end{equation}
Again defining $w=b^{1/2}u$ as a new variable and $\gamma=ab^{-1/2}$ another 
constant quantity, we have from the above equation 
\begin{equation}
\frac{d^2X}{dw^2}+\left (-\frac{1}{4}+\frac{\gamma}{w}\right )X=0
\end{equation}
On comparing this differential equation with that satisfied by the Whittaker function $M_{k,\mu}(x)$
\cite{R8,R9},
given by
\begin{equation}
\frac{d^2}{dx^2}M_{k,\mu}(x)+\left (-\frac{1}{4} +\frac{k}{x}+\frac{\frac{1}{4}-\mu^2}{x^2} \right ) M_{k,\mu}(x)=0
\end{equation}
we have $X(x)\equiv M_{k,\mu}(x)$, for $x=w$, $\mu=1/2$ and $k=\gamma$. Then one can write down the solution in the explicit form as
\begin{equation}
X(w)=M_{\gamma,\frac{1}{2}}(w)=\exp\left (-\frac{w}{2}\right )w
M(1-\gamma,2,w)
\end{equation}
where $M(a,c,x)=~ _1F_1(a;c;x)$, the Confluent Hypergeometric
function. From the properties  of special functions,
clearly, $M(1-\gamma,2,w)$ will be a polynomial and becomes zero for
$w\longrightarrow \infty$, if the parameter
$1-\gamma$ is
$0$ or a negative integer, i.e., $\gamma=n$, for $n=1,2,3 ......$,
the positive integers. This is the physical condition to have a
bounded wave function along $x$-direction.
Under such restricted situation, the solution
can also be expressed in terms of Associated Laguerre function. This alternative form of  wave function is then given by
\begin{equation}
X(w)=\exp\left (-\frac{w}{2}\right )w L_{\gamma-1}^1(w)
\end{equation}
The parameter $\gamma$ is again have to be non-zero positive integer. It can very easily be verified that
the form of eqn(15) is exactly identical with the equation for an 
one-dimensional hydrogen atom \cite{R11,R12}.
Then it is just a matter of simple algebra to show from the
expression for $\gamma$ that this restriction gives 
quantized form  of energy of the particle in an
uniformly accelerated frame. Using the expressions for $a$, $b$ and $E_\perp$, it is straight forward to
show that the quantized form of energy of the particle is given by 
\begin{equation}
E_n=n\hbar\frac{\alpha}{c} =n\hbar \omega ~~{\rm(say)}
\end{equation}
with 
\begin{equation}
\omega=\frac{\alpha}{c}
\end{equation}
where we have put $2^{1/2}\approx 1$ for aesthetic sense and neglected 
$p_y$ and $p_z$ for a purely one-dimensional condition. Then accordingly $E_\perp=m_0c^2$, the rest mass energy of
the particle. Therefore it is quite
surprising result. The differential equation satisfied 
by the particle is exactly identical with
the equation for an one-dimensional hydrogen atom, whereas 
the quantized energy levels are exactly look like that of the energy 
levels for an
one-dimensional quantum harmonic oscillator. 

Therefore we may conclude, by saying that in an uniformly accelerated
frame or in Rindler space, the Schr$\ddot{\rm{o}}$dinger equation for a
particle reduces to the identical form of differential equation satisfied by an one-dimensional hydrogen atom.
Further, the restriction imposed on the solution, to make it physically 
acceptable, gives the quantized energy levels.
The energy levels are found to be exactly identical with that of one-dimensional quantum harmonic oscillator. The
energy levels vary linearly with the quantum number $n$, instead of $1/n^2$, where the last one is the case for an
one-dimensional hydrogen atom. The
energy levels are observed to be independent of particle rest mass. 
It depends only on the acceleration $\alpha$ of the frame.
It is evident from eqn.(19) that $\omega \longrightarrow 0$ as $\alpha \longrightarrow 0$.
Further, unlike the one-dimensional quantum mechanical 
harmonic oscillator, the minimum energy of the particle or the ground state energy for a
given $\alpha$ is
$\hbar \omega$ for $n=1$, i.e., there is no zero point energy. It is
also obvious from the analysis that the energy levels are produced by the
uniform gravitational field or the constant acceleration of the frame. Therefore if any transition takes place from some higher to lower energy
levels, the emitted energy will not be of any kind of conventional or known type quanta.
We call it as the dark quanta or the cosmic phonon and the dark wave for its classical counter part. In the case of excitation to some
higher energy levels the absorbed energy must also be in the form of dark quanta. 

Now from eqn.(20) one can also write $\lambda
\alpha={\rm{constant}}$, which is the gravitational form of Wien's
displacement law  (for the usual black body case it is $\lambda
T={\rm{constant}}$, where $T$ is the temperature of the black body
system). To elaborate this point a little more, we assume that if a large
number of cosmic
phonons are created in the very early universe, 
within a region where the gravitational field is uniform, 
even before the epoch
when the matter and radiation are not decoupled, 
then as the universe expands, since the gravitational field
decreases, the wave length of this non-thermal cosmic phonon 
field will increase.
Therefore, the gravitational field $\alpha$ plays the role of $T$ for
a thermal field in equilibrium, e.g., the CMBR. Therefore based on
our model calculation we may assume that the non-thermal phonon
field, which may also be assumed to be some kind of neutral scalar field
and the thermal field CMBR exist side by side. Whether they can
interact with each other is a matter of further investigation.
Further, since the energy levels are created in presence of
background uniform gravitational field, the concept of spin of the
emitted or absorbed phonons are not considered here. Again defining
some kind of refractive index $\mu \propto \alpha$, we have $\mu
\lambda ={\rm{constant}}$. Since the gravitational field at the
present epoch is low enough, assuming Newtonian gravity, we may write 
\begin{equation}
\alpha \propto \frac{1}{x_l}
\end{equation}
where $x_l$ is the position of the local rest frame, then
\begin{equation}
\lambda\propto x_l^2
\end{equation}
Therefore the wavelength decreases as the square of some length
parameter.
Further, it is quite possible that a large number of such cosmic
phonons might have produce at the proximity of supper massive black
holes present at
the centre of the galaxies. However, in this case since there is no
expansion, these cosmic phonons or the dark quanta or the classical 
counter part, the 
dark waves remain confined at the vicinity of massive black holes. Therefore it is quite likely that these waves
or quanta could be treated as one of the viable candidates for dark matter.
Of course at present we can not give any experimental technique to
detect these cosmic phonons as dark matter.

In some separate analysis \cite{R13}, keeping only the linear term in $x$ in the expansion of $(1+\alpha x/c^2)^{-1}$, we have noticed that the
Schr$\ddot{\rm{o}}$dinger equation can be reduced to a differential form, which is identical with Fowler-Nordheim
equation \cite{R14} for the field emission or cold emission of electrons from a metal surface under the action of a
strong electric field applied near the metal surface. In that analysis we have observed, that particle production
near the event horizon of a black hole, in the non-relativistic approximation is almost identical with the phenomena
of Fowler-Nordheim  emission.

Further, keeping up to the quadratic term in $x$, we have also got an analytical solution for the
Schr$\ddot{\rm{o}}$dinger equation, which is the parabolic cylindrical function.  However, it has no physical
significance.

In our final conclusion, we would like to mention that further study is needed to know the spin, parity etc. for the
emitted or absorbed dark quanta if they really exist.

\end{document}